\documentclass[preprint,showpacs,preprintnumbers,amsmath,amssymb,superscriptaddress,prb]{revtex4}

\usepackage{graphicx}
\usepackage{dcolumn}
\usepackage{bm}

\begin{document}


\title{Gigantic terahertz magnetochromism via electromagnons in hexaferrite magnet Ba$_2$Mg$_2$Fe$_{12}$O$_{22}$}

\author{N. Kida}
\altaffiliation{Present address: Department of Advanced Materials Science, The University of Tokyo, 5-1-5 Kashiwanoha, Chiba 277-8561, Japan}
\affiliation{Multiferroics Project (MF), ERATO, Japan Science and Technology Agency (JST), c/o Department of Applied Physics, The University of Tokyo, 7-3-1 Hongo, Bunkyo-ku, Tokyo 113-8656, Japan}

\author{S. Kumakura}
\affiliation{Multiferroics Project (MF), ERATO, Japan Science and Technology Agency (JST), c/o Department of Applied Physics, The University of Tokyo, 7-3-1 Hongo, Bunkyo-ku, Tokyo 113-8656, Japan}
\affiliation{Department of Applied Physics, The University of Tokyo, 7-3-1 Hongo, Bunkyo-ku, Tokyo 113-8656, Japan}

\author{S. Ishiwata}
\altaffiliation{Present Address: Quantum-Phase Electronics Center (QPEC), The University of Tokyo, 7-3-1 Hongo, Bunkyo-ku, Tokyo 113-8656, Japan}
\affiliation{Cross-Correlated Materials Research Group (CMRG) and Correlated Electron Research Group (CERG), ASI, RIKEN, 2-1 Hirosawa, Wako, 351-0198, Japan}

\author{Y. Taguchi}
\affiliation{Cross-Correlated Materials Research Group (CMRG) and Correlated Electron Research Group (CERG), ASI, RIKEN, 2-1 Hirosawa, Wako, 351-0198, Japan}

\author{Y. Tokura}
\affiliation{Multiferroics Project (MF), ERATO, Japan Science and Technology Agency (JST), c/o Department of Applied Physics, The University of Tokyo, 7-3-1 Hongo, Bunkyo-ku, Tokyo 113-8656, Japan}
\affiliation{Department of Applied Physics, The University of Tokyo, 7-3-1 Hongo, Bunkyo-ku, Tokyo 113-8656, Japan}
\affiliation{Cross-Correlated Materials Research Group (CMRG) and Correlated Electron Research Group (CERG), ASI, RIKEN, 2-1 Hirosawa, Wako, 351-0198, Japan}

\date{\today}

\begin{abstract} 
Effects of temperature (6--225 K) and magnetic field (0--7 T) on the low-energy (1.2--5 meV) electrodynamics of the electromagnon, the magnetic resonance driven by the light electric field, have been investigated for a hexaferrite magnet Ba$_2$Mg$_2$Fe$_{12}$O$_{22}$ by using terahertz time-domain spectroscopy. We find the gigantic terahertz magnetochromism via electromagnons; the magnetochromic change, as defined by the difference of the absorption intensity with and without magnetic field, exceeds $500$\% even at 0.6 T. The results arise from the fact that the spectral intensity of the electromagnon critically depends on the magnetic structure. With changing the conical spin structures in terms of the conical angle $\theta$ from the proper screw ($\theta=0^\circ$) to the ferrimagnetic ($\theta=90^\circ$) through the conical spin-ordered phases ($0^\circ<\theta<90^\circ$) by external magnetic fields, we identify the maximal magnetochromism around $\theta\approx45^\circ$. On the contrary, there is no remarkable signature of the electromagnon in the proper screw and spin-collinear (ferrimagnetic) phases, clearly indicating the important role of the conical spin order to produce the magnetically-controllable electromagnons. The possible origin of this electromagnon is argued in terms of the exchange-striction mechanism.
\end{abstract}


\maketitle

\section{Introduction}
The control of light signal with the magnetic state or conversely the light sensing of the magnetic state is a key issue in the contemporary information technology, as realized in many magneto-optical devices and elements.  Most of the magneto-optical functions exploit the change of light polarization with the magnetic state, usually generating the light polarization rotation (Kerr or Faraday rotation).  On the other hand,  the magnetic-field change of optical absorption itself, i.e., magnetochromism,  is usually very small in spite of the potential demands in spin-electronic science and technology.  This is reasonable since the electronic resonance  is usually tied to the magnetic state only through spin-orbit interaction in the magnetic medium.  Some exceptions are found for the materials endowed with inherent strong charge-spin coupling and magnetically controllable competing orders, such as spin-crossover complexes \cite{ABousseksou}, colossal-magnetoresistive oxides \cite{YOkimoto}, and low-dimensional molecular magnets \cite{JLWhite}. Even for these materials, however, a high magnetic field of several tesla is necessary to cause the gigantic magnetochromism. 

A more promising candidate to host the gigantic magnetochromism is the multiferroic compound, where the electricity and magnetism is inherently coupled \cite{CheongREV, YTokuraREV}. 
Such a coupling is recently evidenced for a variety of oxide compounds, as typified by the magnetic field control of ferroelectricity in TbMnO$_3$ \cite{TKimura_2003} and by the electric field control of the magnetization in GdFeO$_3$ \cite{YTokunaga_2009}. 
As a consequence of such a strong magnetoelectric response, a number of multiferroics may have the potential to show the magnetic resonance in the dielectric constant $(\epsilon)$ spectrum, in addition to (or rather than) the magnetic permeability spectrum.
This collective mode, now termed electromagnon, becomes electrically active and thus can be excited by the light $E$ vector.
This is contrasted by the case of the conventional magnetic resonance (the $k=0$ magnon) driven by the light $H$ vector. 
Indeed, recent optical experiments at terahertz frequencies have revealed the emergence of the electromagnon in a family of multiferroics \cite{PimenovREV,NKidaREV}; $R$MnO$_3$ ($R=$ rare earth ions) \cite{APimenov1}, $R$Mn$_2$O$_5$ \cite{ABSushkov}, Ba$_2$Mg$_2$Fe$_{12}$O$_{22}$ \cite{NKida_Hexaferrite}, CuFe$_{1-x}$Ga$_x$O$_2$ \cite{SSeki_CuFeO2}, and Ba$_2$CoGe$_2$O$_7$ \cite{IKezsmarki}. As the origin of the electromagnons observed in antiferromagnets with non-collinear spin configurations such as $R$MnO$_3$, the exchange-striction mechanism is proposed \cite{RAguilar,SMiyahara}: The non-collinear spins ($S_i$ and $S_j$ on the neighboring $i$-th and $j$-th sites) can produce the local polarization $\Delta P_{ij}$ in response to $E$ component of light, as given by $\Delta P_{ij}\propto\Delta S_i\cdot S_j$. In fact, the spectral shape of the electromagnon in $R$MnO$_3$ can be reproduced by this mechanism \cite{MMochizuki_EM}. As only the scalar product $S_i\cdot S_j$ on the underlying lattice is relevant, the electromagnon should be insensitive to the direction of the spiral spin plane, as already confirmed by the experiments applying the magnetic field on DyMnO$_3$ \cite{PRB2008_Kida}.

\begin{figure}[t]
\begin{center}
\includegraphics{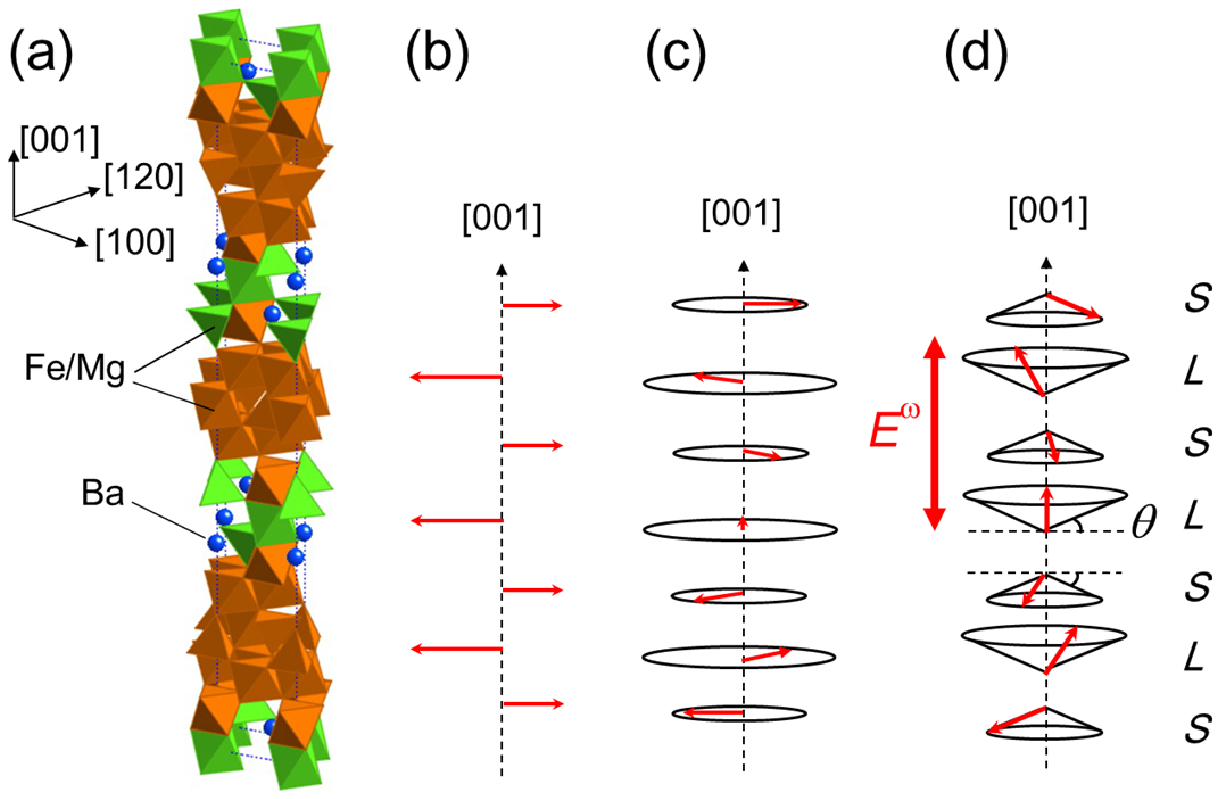}
\end{center}
\caption{(color online) Crystal and spin structures of Ba$_2$Mg$_2$Fe$_{12}$O$_{22}$ in zero magnetic field. Schematic illustrations of (a) hexagonal structure and spin structures in (b) ferrimagnetic, (c) proper screw, and (d) longitudinal conical spin ordered phases. The conical angle $\theta$ is defined in (d). The unit cell can be conveniently divided by two sublattice blocks; spinel $S$ and hexagonal $L$ sublattice blocks along [001]. In the conical spin ordered phase shown in (d), electromagnon becomes active when the light $E$ vector is set parallel to [001].}\label{figCrstal}
\end{figure}

Among several magnetoelectrics to host electromagnons, a hexaferrite, Ba$_2$Mg$_2$Fe$_{12}$O$_{22}$ investigated here, occupies a unique position. Recent combined studies of terahertz time-domain spectroscopy and inelastic neutron scattering have revealed the emergence of the electromagnon around 0.7 THz (corresponding to 2.8 meV in photon energy) \cite{NKida_Hexaferrite}; the electromagnon becomes active in the conical spin ordered phase when the light $E$ vector is set parallel to the propagation vector along [001] [see Fig. \ref{figCrstal}(d)].  

At room temperature, Ba$_2$Mg$_2$Fe$_{12}$O$_{22}$ shows  the ferrimagnetic order composed of two magnetic sublattice blocks, $L$ and $S$ [Fig. \ref{figCrstal}(b)] with the opposite directions of the in-plane magnetizations. Below 195 K, it becomes the proper screw structure, in which the in-plane $L$ and $S$ spins rotate in proceeding along  [001] [Fig. \ref{figCrstal}(c)] \cite{NMomozawa_JPSJ}. Finally, below 50 K, it forms the longitudinal conical spin structure as the in-plane $L$ and $S$ spins decline toward [001] with the conical angle $\theta$ [Fig. \ref{figCrstal}(d)]. Noticeably, in magnetic fields along [001],  Ba$_2$Mg$_2$Fe$_{12}$O$_{22}$ yields the large saturation magnetic moment ($\sim8$ $\mu_B$ per f.u.; f.u., formula unit). This enables us to easily modify the electromagnon inherent to the conical spin phase [see the inset of Fig. \ref{figE001PD}(a)], as compared to other multiferroics with the antiferromagnetic order (no net spontaneous magnetization). Another remarkable characteristic of Ba$_2$Mg$_2$Fe$_{12}$O$_{22}$ is the emergence of the ferroelectricity by controlling the spin cone with magnetic field \cite{SIshiwata_Science, KTaniguchi_APEX}: Transverse conical structure, as realized by declining the cone axis toward [100] [see the inset of Fig. \ref{figE100PD}(a)], can produce the ferroelectricity. In such a transverse spiral or conical magnet, the spontaneous polarization stems from the spin-current \cite{HKatsura,Mostovoy} or inverse Dzyaloshinskii-Moriya mechanism \cite{Dagotto}, as expressed by $\vec{P}=\sum_{\stackrel{<ij>}{}}a_{ij}\vec{e}_{ij}\times (\vec{S_i}\times\vec{S_j})$; here $\vec{S_i}$, $\vec{e_{ij}}$, $a_{ij}$ are the spin at $i$-th site, the unit vector connecting $i$- and $j$-th sites, and the coupling constant, respectively. This scenario is well proved also for this compound by the detailed analysis of the magnetic structure \cite{PRB2010_Ishiwata}. The application of external magnetic field is highly efficient for control of $\theta$; $\theta$ can be changed from 90$^\circ$ to 0$^\circ$. Therefore, the transverse conical structure, where the screw axis is perpendicular to the propagation vector, can host the ferroelectricity along [120]. According to recent neutron scattering experiments, $\theta$ was found to increase from nearly 0$^\circ$ at 50 K to 20$^\circ$ at 25 K \cite{PRB2010_Ishiwata}. From the viewpoint of the magnetochromism, the magnetically-flexible conical spin structure is expected to produce the dramatic modification of the electromagnon spectrum.

For this purpose, we perform the terahertz time-domain spectroscopy and fully map out the spectrum of the electromagnon in a variety of the spin-ordered phases of Ba$_2$Mg$_2$Fe$_{12}$O$_{22}$ by tuning the magnetic field (up to 7 T).  The preliminary experiments in the zero and low magnetic field (up to 0.2 T) have identified the existence of electromagnon \cite{NKida_Hexaferrite}. Here we reveal the gigantic magnetochromism via electromagnons at terahertz frequencies with the fine control of the conical spin order. 
 Furthermore, we could determine the magnetoelectric phase diagram of Ba$_2$Mg$_2$Fe$_{12}$O$_{22}$ by conversely using the electromagnon spectra as the probe. 

The format of this paper is as follows: In Sec. II we describe the experimental setup for terahertz time-domain spectroscopy combined with a split-type superconducting magnet. Section III is devoted to the systematic optical investigations of the electromagnons in a variety of spin-ordered phases tuned by magnetic fields (from the proper screw to the ferrimagnetic through the conical spin-ordered phases), in which we reveal the important role of the conical spin order to produce the electromagnon activity  by the measurements in paraelectric longitudinal (Sec. III A) and ferroelectric transverse (Sec. III B) conical spin-ordered phases. On the basis of the results of Sec. III, we discuss in Sec. IV the possible origin of the electromagnon activity in terms of the symmetric-exchange mechanism. The conclusion is given in Sec. V.

\section{Methods}
The single-crystalline samples were prepared by a flux method as described in Ref. \onlinecite{SIshiwata_Science}, and their magnetic and ferroelectric characteristics were all consistent with previous results. The obtained samples were mechanically polished to the thickness of 70-- 300 $\mu$m  (thickness of the sample is different in the measurements at respective temperatures so as to gain the signal-to-noise ratio). 

\begin{figure}[t]
\begin{center}
\includegraphics{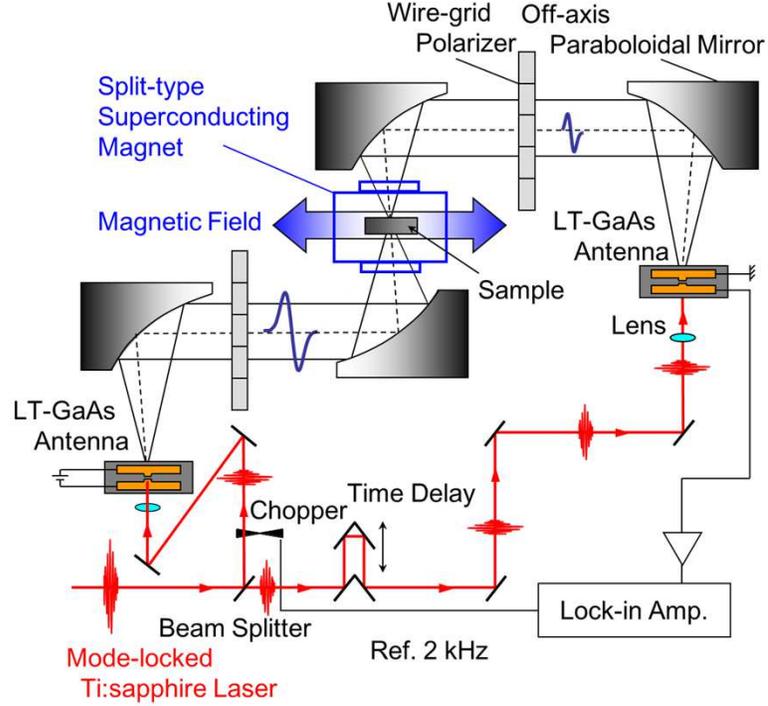}
\end{center}
\caption{(color online) Schematic illustration of experimental setup for terahertz time-domain spectroscopy combined with split-type superconducting magnet. We employed the photoconducting sampling technique with low-temperature-grown GaAs (LT-GaAs) as terahertz emitter and detector. The magnetic field was applied up to 7 T in Voigt geometry to prevent the conventional magneto-optical effect.}\label{figTHzsetup}
\end{figure}

We used the terahertz time-domain spectroscopy in a transmission geometry to directly estimate the complex dielectric constant spectra. Our experimental scheme using the ZnTe crystal as a terahertz emitter was reported in Ref. \onlinecite{PRB2008_Kida}. In the present experiments, to further access the low-energy electrodynamics of the electromagnon, especially down to 0.3 THz (1.2 meV in photon energy), we used the photoswitching sampling technique with a low-temperature-grown GaAs (LT-GaAs) as a terahertz emitter, as schematically shown in Fig. \ref{figTHzsetup}. The femtosecond laser pulses delivered from the mode-locked Ti:sapphire laser (center wavelength of 800 nm; pulse width of 100 fs; repetition rate of 82 MHz) were impinged on the photoswitching device made on the LT-GaAs with dipole-antenna. The radiated terahertz pulse was collimated by a pair of off-axis paraboloidal mirrors and focused on the another photoswitching device made on the LT-GaAs; the induced photocurrent was monitored in time-domain by changing the arrival time of the trigger femtosecond laser pulse. In this setup, we inserted the split-type superconducting magnet (up to 7 T) in between off-axis paraboloidal mirrors. To eliminate the contribution of the conventional magneto-optical effects, we adopted the Voigt geometry, where the light $k$ vector is perpendicular to the direction of the magnetic field. The polarization of the radiated terahertz pulse was carefully tuned by a wire-grid polarizer.

With use of the fast Fourier transformation of the measured terahertz pulse in time domain, we obtained the amplitude and phase spectra with and without the sample. From the obtained complex transmission, we numerically estimated the complex refractive constants $n(\omega)=\sqrt{\epsilon(\omega)\mu(\omega)}$, where $\epsilon$ and $\mu$ are complex dielectric constants and complex magnetic permeability, respectively. As we previously confirmed on the basis of the measurements of the light polarization dependence (both the electric-field and magnetic-field components of light) using crystal plates with different crystal orientations\cite{NKida_Hexaferrite}, the contribution of $\mu(\omega)$ to $n(\omega)$ was negligible in the measured frequency range, namely $\mu(\omega)\approx 1$. In this paper, therefore, we used $\epsilon(\omega)$ as a quantity. Further details of our estimation procedure were reported in Ref. \onlinecite{PRB2008_Kida}.

\section{Results}
\subsection{Effect of magnetic fields on electromagnons in the paraelectric phase}

\begin{figure}[b]
\begin{center}
\includegraphics{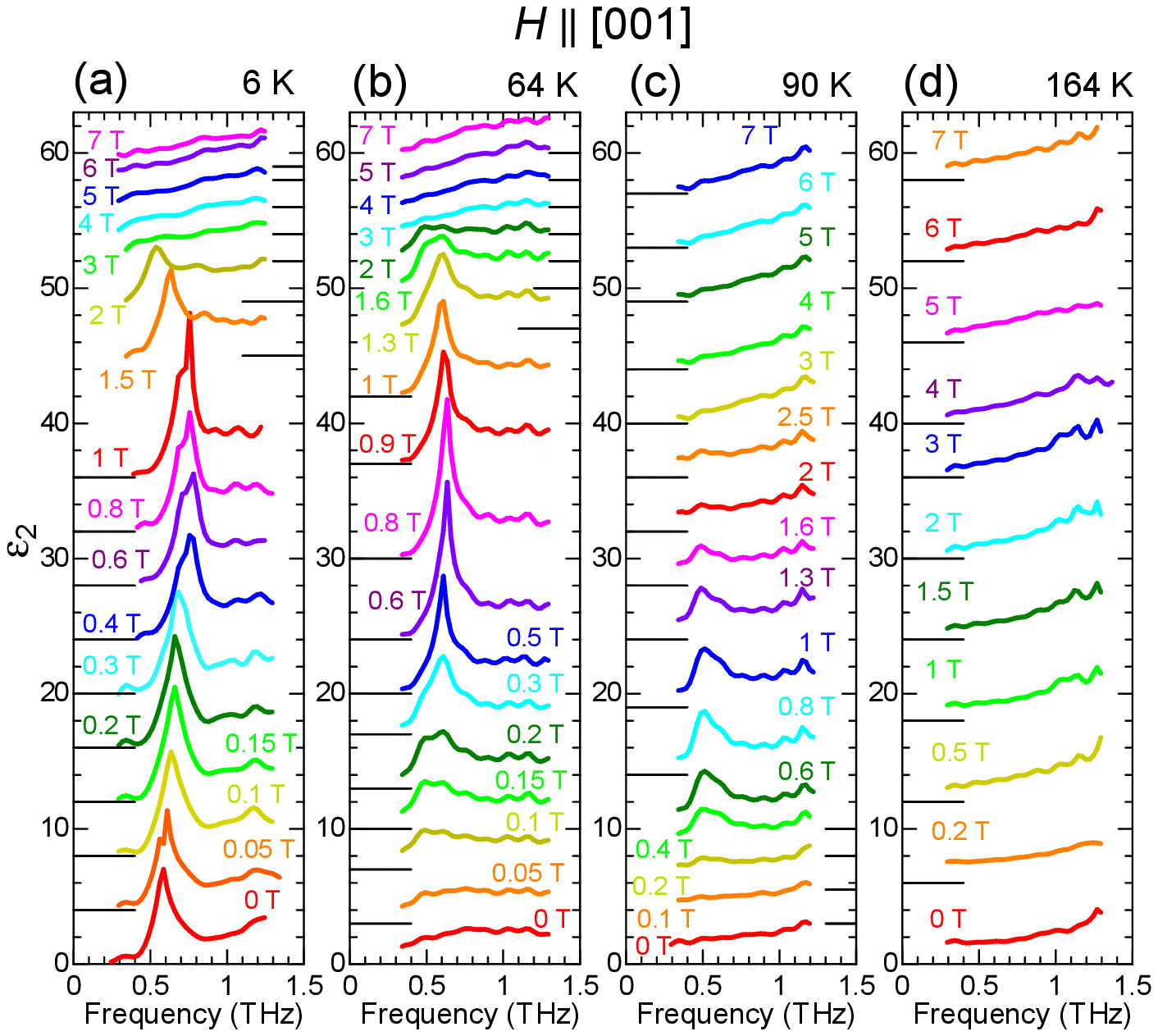}
\end{center}
\caption{(color online) Gigantic magnetochromism via electromagnons in the paraelectric phases of Ba$_2$Mg$_2$Fe$_{12}$O$_{22}$. Imaginary part of the dielectric constant spectra $(\epsilon_2)$ in magnetic fields up to 7 T, measured at (a) 6 K originally in the longitudinal conical spin ordered phase, at (b) 64 K near the conical to proper screw transition temperature, at (c) 90 K and at (d) 164 K originally in the proper screw phase. Respective $\epsilon_2$ spectra are arbitrarily off-set for clarity; the off-set revel is indicated by the horizontal solid line. The light $E$-vector was set parallel to [001] and the magnetic field $H$ was applied along [001] in Voigt geometry.}\label{figE001spectral}
\end{figure}

First, we examine the temperature evolution of the electromagnon in zero magnetic field. The light $E$ vector was set parallel to [001] to induce the electromagnon. In the proper screw phase in the temperature range of 190--50 K, there is no signature of electromagnons in the imaginary part of the dielectric constant $(\epsilon_2)$ spectra, as shown in Figs. \ref{figE001spectral}(b) to \ref{figE001spectral}(d).
Below 50 K, where the longitudinal conical spin order evolves [Fig. \ref{figCrstal}(d)], the sharp resonance can be discerned in $\epsilon_2$ spectrum and its peak magnitude reaches about 6, as exemplified by the $\epsilon_2$ spectrum at 6 K [Fig. \ref{figE001spectral}(a)]. 
This absorption was assigned to the electromagnon, inherent to the longitudinal conical spin ordered phase, on the basis of the combined measurements of the complete set of the light-polarization dependence as well as of the magnetic excitation spectrum with use of the inelastic neutron scattering \cite{NKida_Hexaferrite}.

\begin{figure}
\begin{center}
\includegraphics{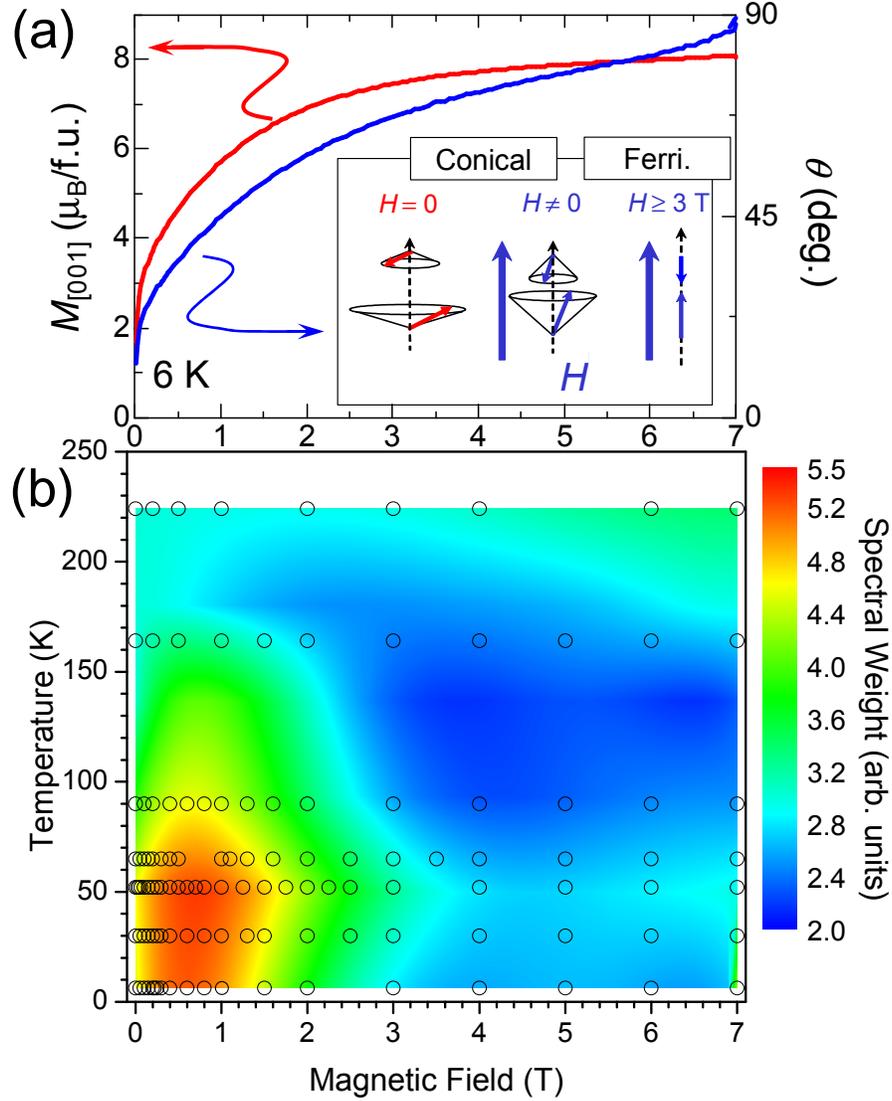}
\end{center}
\caption{(color online) Magnetic phase diagram of Ba$_2$Mg$_2$Fe$_{12}$O$_{22}$, as determined by the signature of the electromagnon. The magnetic field $H$ was applied along [001] to finely control the conical angle $\theta$ of the longitudinal conical structure, as schematically shown in the inset of (a). (a) Magnetization along [001] $M_{[001]}$ and estimate $\theta$ as a function of magnetic field, measured at 6 K. (b) Contour map of the spectral weight (SW) of the electromagnon in terms of temperature and magnetic field. The measured points are indicated by open circles. The magnitude of SW (see text for the definition) is shown with the scale color bar. The unit of SW is arbitrary.}\label{figE001PD}
\end{figure}

 Next, we applied the magnetic field along [001] at 6 K so as to control $\theta$ of the longitudinal conical structure. In the experiments in magnetic fields $H$, we adopted the Voigt geometry to exclude the contribution of the conventional magneto-optical effects such as Faraday rotation. Owing to the ferrimagnetic nature of Ba$_2$Mg$_2$Fe$_{12}$O$_{22}$, the magnetization along [001] increases with magnetic field and yields a large magnetic moment about $\sim8$ $\mu_B$/f.u. at 7 T [Fig. \ref{figE001PD}(a)]. The saturated magnitude of $M$ is consistent with the anticipated ferrimagnetic order, where $L$ and $S$ sublattice blocks were perfectly directed along [001], as schematically shown in the inset of Fig. \ref{figE001PD}(a). Therefore, we could  roughly estimate $\theta$ as a function of magnetic field, as given by $\theta=\sin^{-1}(M/M_s)$ [Fig. \ref{figE001PD}(a)].  At 6 K in zero magnetic field, $\theta$ was estimated to be 20$^\circ$, which is consistent with the measured value by the recent neutron scattering experiments \cite{PRB2010_Ishiwata}. About 3 T, we can induce the ferrimagnetic phase as $\theta$ nearly reaches about 90$^\circ$. In magnetic fields, there is a remarkable change of the electromagnon $\epsilon_2$ spectra [Fig. \ref{figE001spectral}(a)]; this is viewed as the gigantic terahertz magnetochromism. With increasing magnetic field, the electromagnon grows in spectral intensity and reaches the maximum around 1 T, as can be seen in Fig. \ref{figE001spectral}(a). The relative change of the magnitude of $\epsilon_2$ at 0.76 THz with and without magnetic field was estimated to be $\sim$ 300\% in 0.6 T.
 With further increasing magnetic field, the intensity of the electromagnon decreases and the peak position shifts to the lower energy. The electromagnon resonance disappears above 3 T in the ferrimagnetic (all spin collinear) phase.

 Further noticeable magnetochromism can be seen in the $\epsilon_2$ spectra at 64 K on the verge of the transition from the proper 
screw to longitudinal conical spin structures [Fig. \ref{figE001spectral}(b)]. The tendency that the intensity of the electromagnon increases with increasing magnetic field is similar to the case at 6 K [Fig. \ref{figE001spectral}(a)], except for the behavior below 0.5 T. In zero magnetic field, the proper screw state is stable, hence no electromagnon is observed. Even in weak magnetic fields below 0.5 T, we can induce the electromagnon, via the evolution of the longitudinal conical spin order, as identified in the $\epsilon_2$ spectrum around 0.1 T. Therefore, near this temperature, the magnetochromism is strongly enhanced; the relative absorption change yields a gigantic value, $\sim$ 500\% at 0.6 THz even at 0.6 T. With increasing temperature, the value of the relative change tends to decrease, since the application of magnetic field can hardly induce the longitudinal conical spin phase, as exemplified by the $\epsilon_2$ spectra at 90 K [Fig. \ref{figE001spectral}(c)]. Finally, at 164 K, $\epsilon_2$ spectrum is found to be independent of magnetic field ($<7$ T) [Fig. \ref{figE001spectral}(d)]. 

To quantify this trend, we show in Fig. \ref{figE001PD}(b) the contour plot of the spectral weight (SW), as defined by the integration of the optical conductivity (=$\epsilon_0\epsilon_2\omega$, $\epsilon_0$ being a permittivity of vacuum) between 0.4 THz to 1 THz, in the plane of temperature and magnetic field. Open circles represent the measured points. The scale color bar is shown at the right side of the figure; for example, red region indicates the maximum value of the SW. This contour plot clearly reflects the magnetic $(H\parallel [001])$ phase diagram of Ba$_2$Mg$_2$Fe$_{12}$O$_{22}$. In the restricted area of the phase diagram, the longitudinal conical spin ordered phase is identified with the enhanced electromagnon spectral intensity dependent on $\theta$. The present experiments with use of the electromagnon as a probe can provide the full-range phase diagram including the thermal and magnetic-field variations among the proper screw, conical, and collinear structures.

\subsection{Effect of magnetic fields on electromagnons in the ferroelectric phase} 
\begin{figure}
\begin{center}
\includegraphics{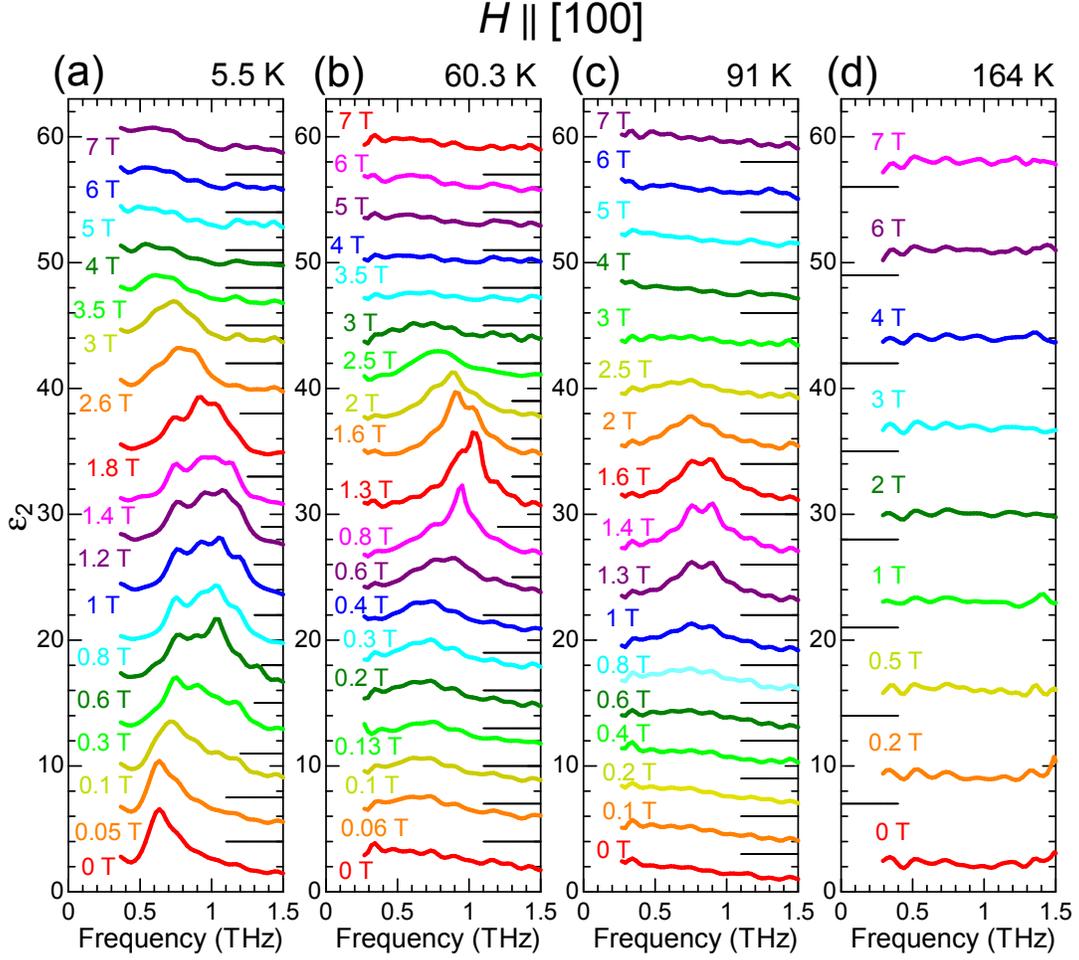}
\end{center}
\caption{(color online) Gigantic magnetochromism via electromagnons in the ferroelectric phase of Ba$_2$Mg$_2$Fe$_{12}$O$_{22}$. Imaginary part of the dielectric constant spectra $(\epsilon_2)$ in magnetic fields up to 7 T measured at (a) 5.5 K originally in the conical spin ordered phase, at (b) 60.3 K near the conical to proper screw transition temperature, at (c) 91 K and at (d) 164 K originally in the proper screw phase. Respective $\epsilon_2$ spectra are arbitrarily off-set for clarity; the off-set revel is indicated by the horizontal solid line. The light $E$-vector was set parallel to [001] and the magnetic field $H$ was applied along [100] in Voigt geometry.}\label{figE100spectra}
\end{figure}

\begin{figure}
\begin{center}
\includegraphics{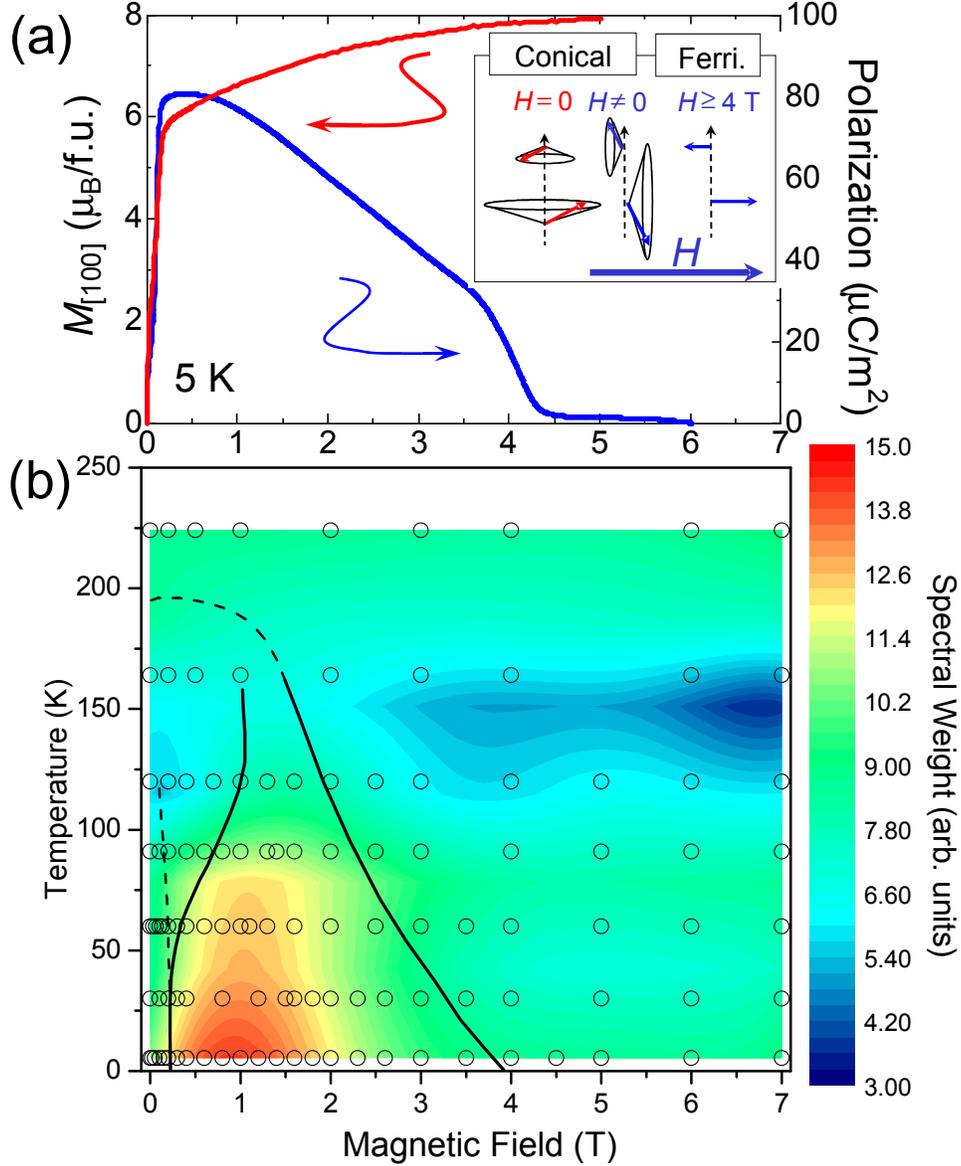}
\end{center}
\caption{(color online) Magnetoelectric phase diagram of Ba$_2$Mg$_2$Fe$_{12}$O$_{22}$, as determined by the signature of the electromagnon. The magnetic field $H$ was applied along [100] to induce the transverse conical spin phase, as schematically shown in the inset of  (a). In this phase, the ferroelectricity emerges along [120], perpendicular to both [100] and [001] (see also Fig. \ref{figCrstal}). (a) Magnetization along [100] $M_{[100]}$ and spontaneous polarization along [120] as a function of magnetic field, measured at 5 K. (b) Contour map of the spectral weight (SW) of the electromagnon in terms of temperature and magnetic field. The measured points are indicated by open circles. The magnitude of SW (see text for the definition) is shown with the scale color bar. The unit of SW is arbitrary. The solid line represents the region, where the transverse conical spin structure was identified by the neutron scattering experiments \cite{PRB2010_Ishiwata}. In this region, the spontaneous polarization emerges, as shown in (a). }\label{figE100PD}
\end{figure}

In Ba$_2$Mg$_2$Fe$_{12}$O$_{22}$,  we can also induce the transverse conical structure by applying the magnetic field along [100].
 In this phase, the ferroelectricity emerges along [120], i.e., along the direction perpendicular to both the magnetic $q$-vector $(\parallel [001])$ and the conical axis $(\parallel [100])$, as is evidenced by the measurements of the spontaneous polarization [Fig. \ref{figE100PD}(a)]\cite{SIshiwata_Science}. Figure \ref{figE100spectra} presents the $\epsilon_2$ spectra at respective temperatures in magnetic fields $(\parallel [100])$ up to 7 T. In zero magnetic field, a signature of the electromagnon at 0.7 THz is discerned only in the longitudinal conical spin phase, e.g., at 5 K [Fig. \ref{figE100spectra}(a)]. When the magnetic field was applied along [100] at 5 K, the magnetization steeply increases around 0.2 T [Fig. \ref{figE100PD}(a)]. Accordingly, the spontaneous polarization emerges and reaches the maximum about 80 $\mu$C/m$^2$. These are all consistent with the transformation from the longitudinal to transverse conical spin orders, which was directly confirmed by the recent polarized neutron scattering experiments \cite{PRB2010_Ishiwata}. The similar magnetochromism is observed in the electromagnon spectra to case of $H\parallel [001]$ (Fig. \ref{figE001spectral}); $\epsilon_2$ increases first with increasing magnetic field and reaches the maximum at around 1 T, and finally, the electromagnon resonance disappears about 3 T, at which the transverse conical spin order transforms to the ferrimagnetic order as evidenced by the reduction of the spontaneous polarization [Fig. \ref{figE100PD}(a)]. Thus, the emergence of the electromagnon is only limited to the conical spin (transverse conical for $H\parallel [100]$, and hence ferroelectric in this case) phase. In turn, we uniquely determine the magnetic phase diagram with use of the electromagnon spectral intensity as a probe. We show in Fig. \ref{figE100PD}(b) the contour plot of the SW (the integrated optical conductivity from 0.3 THz to 1.2 THz) in the plane of temperature and magnetic fields. Thus, we can identify the region, where SW shows the enhancement in the conical spin arrangement, as colored in red and yellow. The solid lines are the guides to indicate the ferroelectric phase in the phase diagram, as inferred from the recent neutron scattering experiments \cite{PRB2010_Ishiwata,HSagayama}, which is consistent with our present data. On the basis of the such mapping of the electromagnon activity in terms of temperature and magnetic field, we confirm again that the conical spin order can act as an effective source of the electromagnons, whose intensity critically depends on $\theta$.

\section{Discussion}

\begin{figure}
\begin{center}
\includegraphics{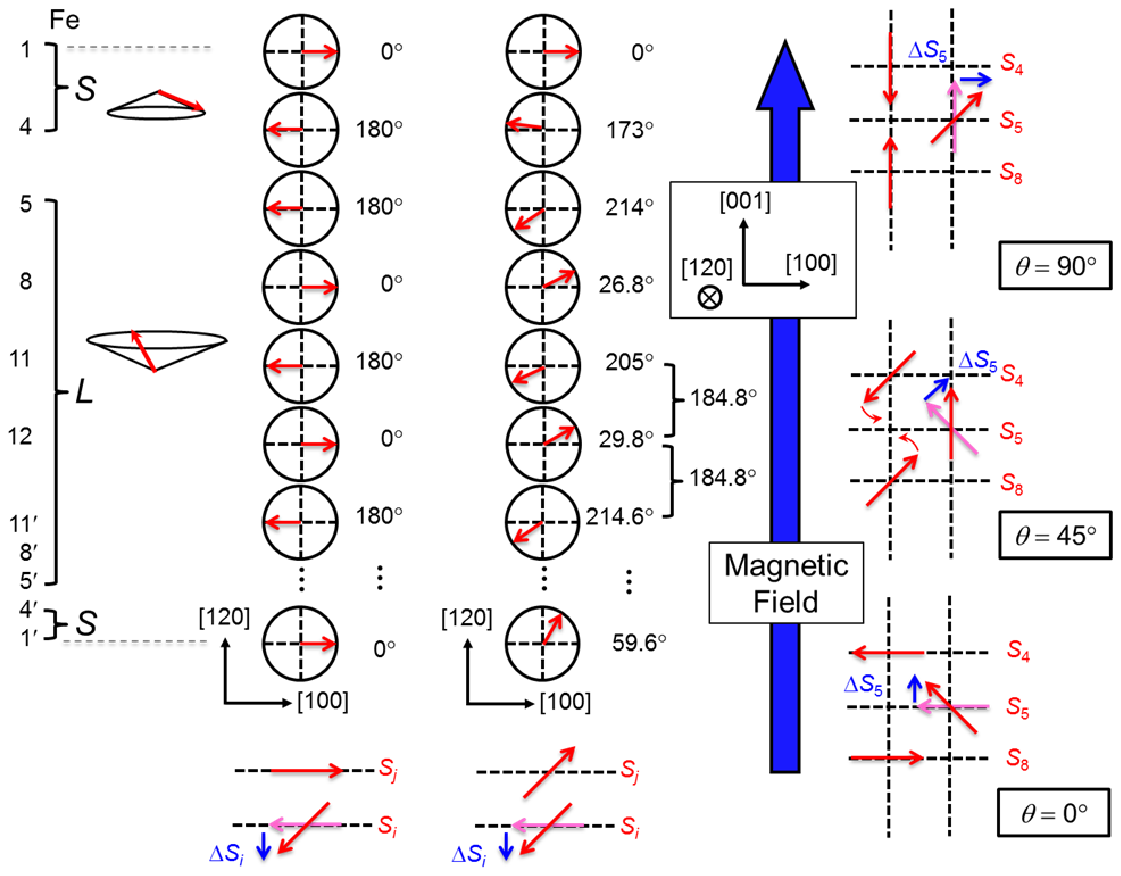}
\end{center}
\caption{(color online) Schematic illustrations of the spin arrangement in Ba$_2$Mg$_2$Fe$_{12}$O$_{22}$. (a) According to the previous neutron scattering experiments \cite{NMomozawa_JPSJ}, $L$ and $S$ sublattice blocks were composed of Fe ions, which are labeled by ``1", ``4", ``5", ``8", ``11", ``12", ``11$^\prime$", ``8$^\prime$", ``5$^\prime$", ``4$^\prime$", and ``1$^\prime$". (b) In the ferrimagnetic phase at 294 K, the spins at respective sites are collinear with a turn angle of 0$^\circ$; the reported data were taken from Ref. \onlinecite{NMomozawa_JPSJ}. (c) In the proper screw phase at 77 K, the spins are non-collinear with a turn angle of 59.6$^\circ$ \cite{NMomozawa_JPSJ}. (d) Possible source of the electromagnons we observed here in the conical spin ordered phase. Within the framework of the symmetric-exchange induced striction, electromagnon becomes active when $\Delta S_i$ is not perpendicular to $S_j$. We focus on the Fe ``4", ``5", and ``8" sites. In the conical phase, the components of the spins projected on to (120) plane can give rise to the electromagnon activity, as schematically shown in (d). In this case, the electromagnon intensity depends on the conical angle $\theta$  and shows the maximum at $\theta$ of $45^\circ$. This can be experimentally realized in magnetic fields. Furthermore, there is no chance to produce the electromagnon activity with $\theta$ of $0^\circ$ or 90$^\circ$. These are all consistent with present experiments.}\label{figDiscussion}
\end{figure}

Although the detailed theoretical study is needed, the most plausible scenario to explain the above behaviors of the electromagnon is to consider the symmetric exchange-striction mechanism. As extensively discussed in $R$MnO$_3$ as the model case \cite{RAguilar,SMiyahara,MMochizuki_EM}, the non-collinear spins with the symmetric (Heisenberg type) exchange interaction $(J S_i\cdot S_j)$ causes the electromagnon activity through the relationship, $\Delta P_{ij}\propto\Delta S_i\cdot S_j$, when the local electric dipole is tied to the $i-j$ bond. According to the detailed magnetic structure analysis of Ba$_2$Mg$_2$Fe$_{12}$O$_{22}$ by neutron scattering \cite{NMomozawa_JPSJ}, $L$ and $S$ sublattice blocks are composed of the stacking of Fe ions along [001], as labeled by ``1", ``4", ``5", ``8", ``11", ``12", ``11$^\prime$", ``8$^\prime$", ``5$^\prime$", ``4$^\prime$", and ``1$^\prime$" [Fig. \ref{figDiscussion}(a)].  Figure \ref{figDiscussion}(b) shows the reported spin arrangement at respective Fe sites within the plane in the ferrimagnetic phase at 294 K \cite{NMomozawa_JPSJ}. In this phase, the spins are all collinear. Therefore, $\Delta S_i$ is perpendicular to $S_j$ and thus $\Delta P_{ij}$ in response to $E^\omega\parallel [001]$ is always zero, as schematically shown in the lower panel of Fig. \ref{figDiscussion}(b). On the other hand, the spins become non-collinear in the proper screw spin phase below 195 K. The spin arrangement within the plane in this phase is schematically shown in Fig. \ref{figDiscussion}(c); the turn angle of $59.6^\circ$ at 77 K was taken from Ref. \onlinecite{NMomozawa_JPSJ}. The electromagnon activity would be possible in this phase [see the lower panel of Fig. \ref{figDiscussion}(c)]. However, this mechanism cannot explain the present experimental results, where there is no remarkable signature of the electromagnon in the proper screw phase [Figs. \ref{figE001PD}(b) and \ref{figE100PD}(b)]. This is partly due to the fact that the in-plane $J$ (scaled with the transition temperature $\sim195$ K) is large compared to the measured photon energy range ($\leq$ 6 meV). In this proper screw spin phase of Ba$_2$Mg$_2$Fe$_{12}$O$_{22}$, $L$ and $S$ spins lie only within the plane [Fig. \ref{figDiscussion}(c)]. On the contrary, spins decline toward [001] in the longitudinal conical ordered phase below 50 K, where the electromagnon shows up. The spin component projected on to (120) plane, emphasized at Fe ``4", ``5", and ``8" sites, is schematically shown in Fig. \ref{figDiscussion}(d); the upper, middle, and lower panels stand for the situations, where $\theta$ is $90^\circ$, $45^\circ$, and 0$^\circ$, respectively. In the case for $0^\circ<\theta<90^\circ$, there emerges another route to $\Delta P_{ij}$ as $\Delta S_i$ is not perpendicular to $S_j$. For example, we can consider $\Delta P$ produced around Fe ``5" site while simply assuming $S_4=(-S^{[100]}, 0, -S^{[001]})$ and $S_8=(S^{[100]}, 0, S^{[001]})$ \cite{Additional}, the situation for $\theta=45^\circ$ is schematically shown in the middle panel of Fig. \ref{figDiscussion}(d). We then obtain non-zero component of $\Delta P$ along [001], being proportional to $2\Delta S_5 S^{[001]}$. In this simple consideration, the intensity of the electromagnon should be scaled with $\sin^2\theta$ and reach the maximum at $\theta=45^\circ$ [middle panel of Fig. \ref{figDiscussion}(d)]. In the longitudinal conical spin phase at 6 K (Fig. \ref{figE001PD}), for example, the observed SW of the electromagnon reaches the maximum around 1 T, which nearly corresponds to $\theta=45^\circ$. In the both ferrimagnetic phases induced by magnetic field along [001] or [100], electromagnons become inactive (Figs. \ref{figE001PD} and \ref{figE100PD}) as all the spins become collinear within the projection of (120) plane [upper panel of Fig. \ref{figDiscussion}(d)], being consistent with this mechanism. Contrary to the origin of the ferroelectricity, i.e. the antisymmetric exchange interaction described by vector spin chirality $S_i\times S_j$, the symmetric exchange interaction $(S_i\cdot S_j)$ acting along [001] is likely the dominant source of the observed electromagnons. Uniquely, such a symmetric exchange interaction depends on $\theta$. Therefore, the magnetic field control of $\theta$ ranging from $0^\circ$ and $90^\circ$ can produce the gigantic terahertz magnetochromism via electromagnons as we observed here.  

\section{Conclusion}
In conclusion, we performed the terahertz time-domain spectroscopy for a variety of the magnetically ordered phases of the hexaferrite Ba$_2$Mg$_2$Fe$_{12}$O$_{22}$, while tuning temperature and magnetic field. The electromagnon appears only in the conical spin ordered phase but irrespective of the longitudinal or transverse conical form, while no electromagnon is observed in the proper screw nor ferrimagnetic phase. The intensity of the electromagnon is found to be highly sensitive to the conical angle tuned by external magnetic fields. Such a sensitivity can provide the unique opportunity to determine the magnetoelectric phase diagram in a wide range of temperature and magnetic field. From the technical viewpoint, we clarified that the electromagnon can induce the gigantic magnetochromism at terahertz frequencies; the estimated relative change of $\epsilon_2$ ($\sim400$\%) is remarkably large even in a magnetic field as low as 0.6 T. The important implication of the present work is that the source of the electromagnon is not limited to multiferroics; electromagnon resonance emerges irrespective of the presence of the ferroelectricity. There are a variety of non-collinear magnets at room temperature, which are candidates to potentially host the gigantic magnetochromism via electromagnons.

\begin{acknowledgments}

We thank S. Miyahara and T. Arima for enlightening discussion.
This work was partly supported by Grants-In-Aid for Scientific Research (Grant No. 20340086 and 2010458) from the MEXT of Japan, and by FIRST Program by JSPS.

\end{acknowledgments}

\clearpage

\clearpage

\clearpage

\clearpage

\clearpage

\begin{thebibliography}{h}
\bibitem{Additional} $S_8$ and $S_4$ belong to $L$ and $S$ sublattice blocks, respectively. Therefore, to be precise, its magnitude should be different. This also produces the non-zero component along [100]. Even in this case, the present discussion is valid in the context that the non-zero component along [001] emerges only in the conical spin ordered phase.
\end{thebibliography}
\end{document}